# TECHNICAL REPORT

ON

A CONTEXT-BASED NUMERICAL FORMAT PREDICTION FOR A TEXT-TO-SPEECH SYSTEM

STUDENT NAME: YASER DARWESH, LIT WEI WERN
w40w@hotmail.com, weiwern92@hotmail.com

SUPERVISOR: ASSOCIATE PROFESSOR DR. MUMTAZ BEGUM MUSTAFA
mumtaz@um.edu.my

FACULTY OF COMPUTER SCIENCE AND INFORMATION TECHNOLOGY, UNIVERSITI MALAYA, 50603 KUALA LUMPUR, MALAYSIA.

**ACKNOWLEDGEMENT**

THIS RESEARCH WAS SUPPORTED BY THE UNIVERSITI MALAYA (BKP SPECIAL) RESEARCH GRANT (GRANT. NO.: BKS012-2018)

**Abstract**

Many of the existing TTS systems cannot accurately synthesize text containing a variety of numerical formats, resulting in reduced intelligibility of the synthesized speech. This research aims to develop a numerical format classifier that can classify six types of numeric contexts. Experiments were carried out using the proposed context-based feature extraction technique, which is focused on extracting keywords, punctuation marks, and symbols as the feature of the numbers. Support Vector Machine, K-Nearest Neighbors Linear Discriminant Analysis, and Decision Tree were used as classifiers. We have used the 10-fold cross-validation technique to determine the classification accuracy in terms of recall and precision. It can be found that the proposed solution is better than the existing feature extraction technique with improvement to the classification accuracy by 30% to 37%. The use of the number format classification can increase the intelligibility of the TTS systems.

# 1.0 Introduction

The Text-to-speech (TTS) system has a diverse type of users from people with visual impairments or reading disabilities to non-impaired users. TTS systems has been incorporated into many applications such as read-aloud for webpages (Gonçalves, Rocha, Martins, Branco, & Au-Yong-Oliveira, 2018), document readers (Anderson, 2018), robotics (Hossain, Amin, Islam, Marium-E-Jannat, 2018), and speech to speech translation (Tran, Ahamed Khan, & Sridevi, 2020). TTS development is not only focused on well-resourced languages such as English, Spanish, and Mandarin, but also many under-resourced languages such as Malay, Indonesia, Vietnamese and so on (Aliero, Muhammed, Mustafa, Muhammad, Garba, 2020; Jayakumari, & Jalin, 2019; Jha, Singh, & Dwivedi, 2019; Nthite, & Tsoeu, 2020).

In an ideal situation, there would be an unambiguous relationship between spelling and pronunciation. But in real text, there are non-standard words such as numbers, digit sequences, acronyms, abbreviations, dates, etc, resulting in generated speech with poor intelligibility (San-Segundo, Montero, Giurgiu, Muresan, & King, 2013).

Many speech and language applications require text tokens to be converted from one form to another. For example, in text-to-speech synthesis, one must convert digit sequences (32) into number names (thirty two), and appropriately verbalize date and time expressions (12:47 as twelve forty-seven). The inability of converting non-standard words (NSWs) into normalized words resulted in incorrect speech output that degrades the TTS system's intelligibility (Burkhardt & Reichel, 2016).

The existing TTS systems do not contain a specific number to text converter. Instead, they use auxiliary facilities called "Text-conditioning tools" such as lexical analyzer and semantic analyzer to rewrite the rules and produce the semantics of the numbers (Kardava, Antidze, & Gulua, 2016; Sproat et al., 2001). To produce the correct semantic representation for numbers, it is important to predict its context before being converted into speech (San-Segundo, Montero, Giurgiu, Muresan, & King, 2013). The conversion of numerical formats can be made better if the format of the numbers can be determined prior to the conversion. However, no effort has been made in the existing literature on the prediction of numbers format from the text input.

The overall aim of this study is to automatically classify the numerical format of a text, which will improve the intelligibility of the synthetic speech generated by the TTS systems. This study focuses on classifying the numerical format of the input text for Malay language, which is one of the under-resourced languages (Mustafa & Ainon, 2013; Mustafa, Don, Ainon, Zainuddin, and Knowles, 2014). However, the idea capitalized in this research can be applied to any of the languages.

The rest of the paper is organized as follow: section 2 provides background of the research and the ability of number to text conversion by the existing TTS systems. Section 3 discusses the techniques for predicting the numerical format. Section 4 covers the experimental design of the study. Section 5 presents and discusses the major findings, while section 6 concludes the study.

## 2.0. Research Background

### 2.1 Numbers Normalization

Burkhardt & Reichel (2016) state that both the commercial and open-source TTS systems have a long way to go towards achieving an effective system that can handle all forms of inputs including numbers and units, special characters, and addresses. Swetha & Anuradha (2013) state that TTS systems have to deal with NSWs, which include abbreviations, acronyms, addresses, numbers, and special characters.

Most of the existing TTS systems use lexical analyzer, a pre-designed component that withdraws the indivisible text characters (lexemes) and clustered them into token (Clark & Araki, 2011; Han, Cook, & Baldwin, 2013). However, manual input is required to define the rules for converting numbers, dates, time, and other symbolic representations (Aida-Zade, Ardil, & Sharifova, 2013).

Gorman and Sproat (2016), stated that while users might forgive a TTS system that reads the ambiguous abbreviation such as 'plz' as 'plaza' rather than the intended word 'please', it would be inexcusable for the same system to ever read '72' as 'four hundred seventy two', even if it rendered the vast majority of numbers correctly". The best way for numbers normalization is using machine learning solutions due to the amount of time, effort, and expertise required to produce error-free number grammars (Gorman, & Sproat, 2016).

### 2.2. Name Entity Recognition for Numbers in Malay Language

Akinadé, and Ọdéjọbí, (2014) state that naming number in human languages requires various mathematical and linguistic processes. For example, the number '74' is represented as '70' increased by '4' in English, whereas it is represented as 60 increased by 14 in French (Akinadé, & Ọdẹ́jọbí, 2014).

Numbers exist in a variety of formats, such as date, time, measurement, currency and etc, where these formats can be similar to each other making it a challenge to determine the exact number format. NLP and text normalization processes can be performed more accurately by classifying the context of numbers (Shetake, Patil, & Jadhav, 2014).

Named Entity Recognition (NER) is one of the sub-tasks of NLP which has many applications mainly in natural language understanding, text-to-speech synthesis, information extraction, information retrieval, machine translation, question answering, and etc.

A rule-based NER uses hand constructed rules of grammar and imposes linguistic constraints in the classification of named entities in unstructured textual content as well as documents such as news and academic articles. Rule-based NER systems are useful in NLP discipline but it has several drawbacks such as it is language dependent and rigid.

Alternatively, Machine learning (ML) can be used in NER where the computer can learn from data and information autonomously by using computer algorithms. It can change and improve the algorithms automatically without changing their program every time. ML techniques such as conditional random fields (CRF), Maximum Entropy Markov Model (MEMM), Support Vector Machine (SVM) and Hidden Markov Model (HMM) are commonly used for NER (Gorman, & Sproat, 2016; Morwal, Jahan, & Chopra, 2012). The training phase consists of two main steps, which are the extraction of features and model generation. From the training data, domain knowledge will be used to extract new features that will be utilized by the ML technique. BoW and TF-IDF are some examples of popular text feature extraction. Lodhi, Saunders, Shawe-Taylor, Cristianini, & Watkins (2002) have successfully applied the BoW technique using SVM classifier with a classification accuracy of 97%.

**2.3 Classifications of Number Format**

Several classification techniques have been proposed in the literature that is suitable for classifying the numerical format. However, it is not easy if not impossible to decide which of the existing classification techniques perform better than others (Brownlee, 2015). The difficulties in selecting suitable classifier are generally due to the diverse nature of the problem to be solved (Rekha, Srinivas, & Reddy, 2014; Jagan & Rajagopalan, 2015), which include the data size, and quality. It also depends on the training time, i.e. the amount of time needed to train the model.

However, there are other considerations that must be given when selecting a technique such as the accuracy, the number of parameters, the number of features, and linearity (Rohrer, Franks, & Ericson, 2016). Rekha et al. (2014) provide an excellent review of several classification techniques that can be used in text classification such as Support Vector Machine (SVM), K-Nearest Neighbor (KNN), Linear Discriminant Analysis (LDA), Decision Tree (DT), and Naïve Bayes. These techniques have been used to tackle text classification problems in various domain (Browne & McNicholas, 2012; Korde & Mahender, 2012; Tahir, Kittler, & Bouridane, 2016; Trstenjak, Mikac, & Donko, 2014).

## 3.0 The Context-Based Numerical Format Prediction

There are three stages in the development of the proposed numerical format classification system using ML technique as shown in Fig.1. The first stage is the extraction of the context-based features of numbers from the dataset. Information needed to train the classifiers includes the features and its labels. The second stage is the training of the classifier model, where features of numbers and its labels collected in stage one is used to train the model. The third stage is the classification and evaluation where the trained model will be used to classify test data in order to evaluate its accuracy and efficiency of the trained model.

In this research, the general idea of the context-based technique is that the system extracts the keywords and symbols nearby the numbers. The keywords consist of collective nouns, measurement units, currency code, etc., which are usually positioned next to the detected numbers. On the other hand, the symbol includes punctuation such as currency symbols, commas, full stops, etc.

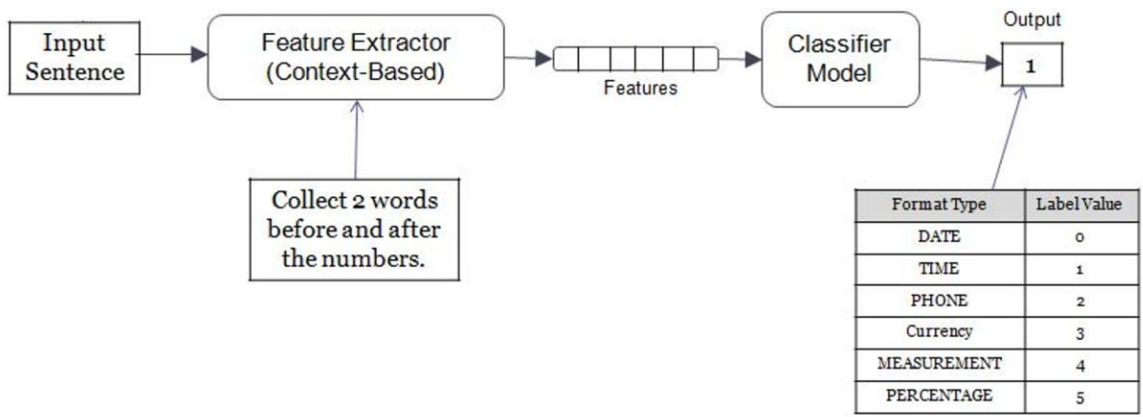

Figure 1: The Context-Based Numerical Format Prediction

The Malay language belongs to one of the western branches of the Austronesian language family, and it is widely spoken by more than 150 million people in Malay-speaking countries such as Malaysia, Indonesia, Singapore, and Southern Thailand (Mustafa et al., 2014).

The format of numbers used in Malay text is most of the time similar to of English language. For instance, the use of full stop (for decimals) and comma (for thousands) is the same for both the English and Malay (Ranaivo-Malançon, 2006). Fig 2 depicts the similarities and differences of number format between of English and Malay. The similarity includes the way percentage and measurement is converted into text. There are more ways the numbers can be converted into text for Malay in term of year, time, and currency than for English.

| Format | English | Malay |
| --- | --- | --- |
| Year | 1924: nineteen twenty four | 1924: Sembilan belas dua puluh empat<br>1924: seribu sembilan ratus dua puluh empat |
| Time | 2.00 PM: Two PM<br>06.00: Zero six hundred hours<br>14.00: Fourteen hundred hours | 2.00 PM: Dua petang<br>06.00: Enam pagi<br>14.00: Dua petang |
| Currency | $ 2: Two dollars<br>$ 0.50: Fifty cents<br>$ 2.50: Two dollars fifty cents | RM 2 : RM dua<br>RM 2: Dua Ringgit<br>RM 0.50: RM Lima puluh sen<br>RM 2.50: Dua Ringgit lima puluh sen<br>RM 2.50: Dua Ringgit setengah |
| Measurement | 5 cm: Five cm<br>5 cm: Five centimeter | 5 cm: Lima cm<br>5 cm: Lima sentimeter |
| Percentage | 5%: five percent | 5%: lima peratus |

Figure 2. The Similarities and Differences of Number Format between English and Malay.

The context-based model is used for text categorization, where the neighboring words to the numbers are used as features for training purposes. The feature extraction program will extract two neighboring words, preposition 1 (1st word before the number), preposition 2 (2nd word before the number), postposition 1 (1st word after the number), and postposition 2 (2nd word after the number). Two words before and after the numbers are collected because most of the keywords in describing the number appear in between this range. Table 1 shows the samples of context-based models for sentences with different number formats.

Table 1: Sample of context-based models

| Sentence 1 | *Mahkamah menetapkan 21 Januari ini untuk sebutan semula kes* (The court set this January 21 for re -mention of the case) |
|---|---|
| Preposition 2 | *Mahkamah* (court) |
| Preposition 1 | *Menetapkan* (set) |
| Post position 1 | *Januari* (January) |
| Post position 2 | *Ini* (this) |

## 4.0 Experiments

Experiments were carried out using the proposed context-based feature extraction technique together with several classifiers. We describe the creation of the dataset and experimental procedures as well as the performance measures. We have compared our proposed technique with the Bag of Word (BoW) techniques as the benchmark.

## 4.1. Data Preparation

A Malay dataset is needed to perform the NER classification, which is the objective of this research. As there is no existing and compatible dataset that can be used to classify numbers in text, a new database with a proper label is built in this research. Several resources have been used in this research to develop the appropriate dataset using online Malay medium local newspaper such as the *Berita Harian, Utusan Online, MalaysiaKini*, and *Harian Metro.*

The dataset contains six categories of number formats which are Calendar Date, Time, Phone Number, Currency, Measurement, and Percentage. We have decomposed each of the categories into sub-categories as shown in Table 3. Punctuation marks such as full stop, comma, colon, hyphen, brackets, and slash in numbers can be used to distinguish number format. For instance, currency symbols such as the Ringgit sign "RM" commonly appear together with the number in a sentence to indicate the number is currency format. Punctuation marks are also used as features in this research to increase the accuracy of the proposed system. The list shown in table 2 is non exhaustive as more format can be added in the future to increase the performance of the system.

Table 2: The sub-categories of each number format

| Number Type | Sub-Categories |
|---|---|
| Calendar Date | xx month |
| | xx month xx |
| | xxxx-xx-xx |
| | xx/xx/xxxx |
| | xxxx-xx |
| | tahun xxxx (year xxxx) |
| Time | xx *pagi/petang/malam/tengah hari* (x am/pm) |
| | x.x *pagi/petang/malam/tengah hari* ( x.x am/pm) |

| Number Type | Sub-Categories |
| --- | --- |
| | xx.xx *pagi/petang/malam/tengah hari* (xx.xx am/pm) |
| | *jam* x.xx *pagi/petang/malam/tengah hari* (at x.xx am/pm) |
| | *jam* x:xx *pagi/petang/malam/tengah hari* (at xx.xx am/pm) |
| | *pukul* x *pagi/petang/malam/tengah hari* (at x am/pm) |
| | xx:xx am/pm |
| Phone Number | xx-xxxxxxx |
| | +xx-xx-xxxxxx |
| Currency | RM xx |
| Measurement | xx (collective nouns, measurement units) |
| Percentage | *Peratus* (percentage) |
| | % |
| | xx *hingga* xx *peratus* (xx to xx percent) |
| | xx – xx *peratus* (xx - xx percent) |

*Note: xx indicate any number

## 4.2. Experimental procedures

### 4.2.1. Benchmark Feature Extraction

To extract the feature from the training data that can be used as a features vector, a matrix was created to carry all the features on dimensions of 5,291 x 1,000. The bag-of-words (BoW) was used as a tool for feature extraction. Before that, it is essential to use the n-gram of the size 1 (Unigram) and then convert each gram (e.g. character/digit) into the uint8 data type. The uint8 data type holds the values between 0-255, where the minimum value is 0 and the maximum value is 255.

Using the number "1500" as an example, when unigram model was applied, the number "1500" will be segmented into 1-gram sequence as follows: {'1', '5', '0', '0'}, where each digit called gram. Then, each gram will be converted into uint8 data type as follows: {'49', '53',

'48', '48'}, where these values represent the feature vector values for the number 1500. The total features found in the bag are 38,396 (with only 55 unique types of features) for the dataset developed in this research.

*4.2.2. Proposed Feature Extraction Technique*

Python programming language is the tool used to implement the context-based features extractor. First of all, the system will collect the sentences available in CSV file, and then locate the number in the sentence by using the python's find function. After locating the number in the sentence, it will split the sentence into separated words to determine the two immediate words before and after the numbers. Table 3 shows the some of the list of the keywords used in this research.

Table 3: The list of the keywords selection in the present study

| Types | Keywords |
|---|---|
| Calendar Date | *januari, februari, mac, april, mei, jun, julai, ogos, september, oktober, november, disember (*January, February, March, April, May, June, July, August, September, October, November, December*)* |
| Time | *am, pm, masa, pukul, jam, minit, saat, pagi, petang, malam, tengah* (am, pm, time, hour, minute, second, morning, evening, night, afternoon) |
| Phone Number | *tel, telefon, talian, menghubungi* (tel, telephone, line, connecting) |
| Currency | ringgit, baht, birr, boliviano, dalasi, deutsche, dinar, dobra, dolar, euro |
| Measurement Units | meter, kilometer, gram, kilogram, ampere, kelvin, mol, kandela, |
| Collective Nouns | *baris, bidang, biji, bilah, blok, bongkah, botol, buah, buku, bungkus, butir, carik, , colek, cubit, cucuk, das, deret, ekor, gelas, gelung, lingkar, ikat, jambak, jambangan, jemput, kajang, kaki, kalung, kandang, kumpulan, kuntum, laras, lembar, longgok, mangkuk, naskhah, orang, puntung, tandan, tangkai, teguk, timbun, tingkat, titik, titis, tongkol, ulas* (row, field, seed, blade, block, |
|            | chunk, bottle, fruit, book, wrap, grain, rip,, poke, pinch, prick, das, row, tail, glass, loop, circle, tie, bunch, vase, pick , awning, legs, necklace, cage, collection, buds, barrel, sheet, pile, bowl, copy, person, butt, bunch, stalk, gulp, pile, level, point, drop, cob, review) |
| Percentage | *Peratus* *(pecent, ppercentage)*                                                                                                                                                                                                                                                                                 |
| Place Value | *puluh, ratus, ribu, juta, bilion, trilion, kuadrilion, kuintilion* (tenth,, hundred, thousand, million, billion, trillion, quadrillion, Quintillion)                                                                                                                                                        |
| Others     | *nilai, bernilai, jumlah, berjumlah, harga, berharga,* seramai (value, worth, amount, amount to, price, valuable, amount to*)*                                                                                                                                                                                   |

The proposed solution is focused on extracting keywords, punctuation marks, and symbols as the feature of the numbers. In this research, there are a total of 103 different features that have been extracted from the 571 labeled data, which included keywords, punctuation marks, and symbols.

*4.2.3. Training and Testing of the Classifiers*

In this research, there are four main classification techniques applied, which are SVM, LDA, KNN, and DT. Besides that, there are different techniques of kernel trick for SVM classifier that was used, such as Polynomial (poly), Linear, and Gaussian (RBF). On the other hand, for KNN, a single nearest neighbor (k=1) was chosen along with three nearest neighbors (k=3).

**4.3. Evaluation method**

Cross-validation proposed in Ekbal and Bandyopadhyay (2008) is used to measure the effectiveness of each classifier. It is a popular method due to its simplicity to understand and its less biased estimate of the model skill. The classification accuracy is measured in term of precision, recall and F-Measure, which are determined using the following formulas.

$$Precision = \frac{True\ Positive}{True\ Positive + False\ Positive} \quad (1)$$

$$Recall = \frac{True\ Positive}{True\ Positive + False\ Negative} \quad (2)$$

$$F - Measure = \frac{2 \times Precision \times Recall}{Precision + Recall} \quad (3)$$

Testing of the classifier was performed using 10-fold cross-validation method, where 90 percent of the dataset was used for training and the remaining 10 percent was used for testing.

## 5.0. Results and Discussion

Table 4 shows the summary of the classification accuracy that comprises the highest accuracy, the mean accuracy and the standard deviation. The highest accuracy refers to the maximum accuracy that a particular classifier, while the mean looks at the overall performance of the classifier in each experiment. On the other hand, the standard deviation is aimed at measuring the consistency of the classifiers in classifying the number formats.

Table 4: The Classification Accuracy of the Classifiers

| Classifier | Highest Accuracy (%) | Mean Accuracy (%) | Standard Deviation (%) |
|---|---|---|---|
| SVMpoly | 96.49 | 92.12 | 4.13 |
| SVMrbf | 89.29 | 78.14 | 7.53 |
| SVMlinear | 100.00 | 93.86 | 3.83 |
| LDA | 94.83 | 90.39 | 3.82 |

| Classifier | Highest Accuracy (%) | Mean Accuracy (%) | Standard Deviation (%) |
|---|---|---|---|
| KNN1 | 96.49 | 91.07 | 3.92 |
| KNN3 | 96.43 | 89.16 | 5.26 |
| DT | 98.25 | 94.37 | 2.96 |

In this research, SVM with linear kernel reported the highest classification accuracy of 100%. DT classifier achieved the highest mean classification accuracy of 94.37%, followed with SVM with linear kernel at 93.86%, and SVM with polynomial kernel classifier with mean accuracy of 92.12%. The mean classification accuracy of LDA, KNN1, KNN3, and SVMrbf is 90.39%, 91.07%, 89.16%, and 78.14% respectively.

The standard deviation shows the variation in the recognition accuracy of each ML. The low value of the standard deviation indicates the consistency of the ML in classifying the various number formats in text. The ML with the lowest standard deviation is the DT, indicating the ability of the DT to consistently generate similar classification accuracy for each of the 10-fold validation.

Table 5 shows the confusion matrix for each number formats. In terms of recall, the time format is the highest, indicating that all of the selected classifiers can classify the time format with the highest accuracy. This may be because the variation to time format is lower than other number formats and is distinct from other forms of number formats. On the other hand, calendar date format has the lowest precision, notably confused with the measurement format. This is because, calendar date and measurement have similar structures of writing such as the involvement of collective nouns. For example, "*suhu hari in kekal pada 35 darjah Celsius*" (the temperature today is fixed at 35 degrees Celsius) and "*tarikh tutup* *peraduan masih kekal pada 25 Julai*" (The closing date for the contest remains on July 25th) has similar collective nouns and thus similar preposition.

In term of precision, phone number format is the highest, while measurement format is the lowest due to similarities in the text structure between measurement, date, and percentage. However, this has little impact on the accuracy of speech generated by the TTS system as the pronunciation of numbers for these formats is the same for Malay language.

Table 5: The confusion matrix for each number formats

| Number Format | Classification by ML | | | | | | Recall (%) |
|---|---|---|---|---|---|---|---|
| | Date | Time | Phone | Currency | Measurement | Percentage | |
| Date | 69 | 0 | 0 | 0 | 15 | 0 | 82.14 |
| Time | 0 | 89 | 0 | 0 | 0 | 0 | 100.00 |
| Phone | 0 | 1 | 9 | 0 | 0 | 0 | 90.00 |
| Currency | 1 | 0 | 0 | 76 | 0 | 0 | 98.71 |
| Measurement | 0 | 0 | 0 | 2 | 68 | 0 | 97.14 |
| Percentage | 0 | 0 | 0 | 1 | 3 | 81 | 95.29 |
| Precision (%) | 98.57 | 98.88 | 100.00 | 96.0 | 79.07 | 100.00 | |

The mean classification accuracy of BoW technique using classifiers such as SVM, DT, LDA, and KNN are 56.96%, 61.69%, 55.91% and 59.74% respectively. The mean classification accuracy for BoW technique is lower than context-based technique as shown in Table 6. The proposed solution outperform the benchmark BoW with improvement to the classification accuracy by 30% to 37%.

Table 6: Mean Classification Accuracy of the Proposed Context-Based and the BoW Technique

| **Mean Classification Accuracy** |
|---|

|  | DT (%) | SVMlinear (%) | KNN1 (%) | LDA (%) |
|---|---|---|---|---|
| Context-Based Technique | 94.37 | 93.86 | 91.07 | 90.39 |
| BoW Technique | 56.96 | 61.69 | 55.91 | 59.74 |

## 6.0. Conclusions

From the experiment, it is clear that some of the number formats are more challenging than others such as the Calendar Dates and Measurement. The difficulty in the prediction lies in the possible variation in the way these formats are presented in the text. For example, Calendar Date in US is written differently from the UK. On top of that, some of the numbers can represent different formats. For example, fraction can also be interpreted as Calendar Date. The use of the number format classification can increase the intelligibility of the TTS systems.

We also found that DT classifier is the most reliable classifiers in classifying the text format in this research. This make sense as DT has the advantage over other techniques such as lower effort for data preparation during pre-processing, not requiring normalization of data as well as scaling of data. Since the proposed solutions looks at very close neighboring words of the numbers in written text, the simplicity of the data for classification makes it very effective for DT in classifying number format from text.

Some of the future works that can be carried out for both the well-resourced as well as other poor resourced languages is the development of a much larger text database that can improve the classification accuracy of the number format for text. A multilingual number format classification will also be very useful for languages that share similar number normalization.

**References**

Aliero, A.A., Muhammed, D., Mustafa, M.B., Muhammad, S.A., Garba, M. (2020). A Cross-Lingual Text-To-Speech System for Hausa using DNN-Based Approach, International Journal of Mechatronics, Electrical and Computer Technology (IJMEC), Vol. 10(35), 4460-4470.

Aida-Zade, K. R., Ardil, C., & Sharifova, A. M. (2013). The main principles of text-to-speech synthesis system. *International Journal of Signal Processing*, *7*(1), 13–19.

Akinadé, O. O., & Ọdéjọbí, Ọ. A. (2014). Computational modelling of Yorùbá numerals in a number-to-text conversion system. *Journal of Language Modelling*, *2*(1), 167-211.

Anderson, T. (2018). E-readers make a Difference for Diverse Readers. *International Journal of Technology in Education and Science*, *2*(1), 40-56.

Browne, R. P., & McNicholas, P. D. (2012). Model-based clustering, classification, and discriminant analysis of data with mixed type. *Journal of Statistical Planning and Inference*, *142*(11), 2976–2984.

Brownlee, J. (2015). Choosing Machine Learning Algorithms: Lessons from Microsoft Azure - Machine Learning Mastery. Retrieved March 18, 2017, from http://machinelearningmastery.com/choosing-machine-learning-algorithms-lessons-from-microsoft-azure/

Burkhardt, F., & Reichel, U. D. (2016). A Taxonomy of Specific Problem Classes in Text-to-Speech Synthesis : Comparing Commercial and Open Source Performance. *In: Proceedings of the Tenth International Conference on Language Resources and Evaluation (LREC 2016). European Language Resources Association (ELRA), Portoroz, Pp. 744-749. ISBN 978-2-9517408-9-1*.

Clark, E., & Araki, K. (2011). Text normalization in social media: Progress, problems and applications for a pre-processing system of casual English. In *Procedia - Social and Behavioral Sciences* (Vol. 27, pp. 2–11).

Ekbal, A., & Bandyopadhyay, S. (2008). Bengali named entity recognition using support vector machine. In *Proceedings of the IJCNLP-08 Workshop on Named Entity Recognition for South and South East Asian Languages*.

Gonçalves, R., Rocha, T., Martins, J., Branco, F., & Au-Yong-Oliveira, M. (2018). Evaluation of e-commerce websites accessibility and usability: an e-commerce platform analysis with the inclusion of blind users. *Universal Access in the Information Society*, *17*(3), 567-583.

Gorman, K., & Sproat, R. (2016). Minimally supervised number normalization. *Transactions of the Association for Computational Linguistics*, *4*, 507-519.

Han, B., Cook, P., & Baldwin, T. (2013). Lexical normalization for social media text. *ACM Transactions on Intelligent Systems and Technology*, *4*(1), 1–27.

Hossain, M. J., Amin, S. M. A., Islam M. S., Marium-E-Jannat. (2018). Development of robotic voice conversion for RIBO using text-to-speech synthesis, 2018 4th International Conference on Electrical Engineering and Information & Communication Technology (iCEEiCT), Dhaka, Bangladesh, 2018, pp. 422-425, doi: 10.1109/CEEICT.2018.8628115.

Jagan, S., & Rajagopalan, S. . (2015). A Survey on Web Usage Mining. *International Research Journal of Engineering and Technology (IRJET)*, *02*(1), 6–12.

Jayakumari, J., and Jalin, A. F. (2019). An Improved Text to Speech Technique for Tamil Language Using Hidden Markov Model, 2019 7th International Conference on Smart Computing & Communications (ICSCC), Sarawak, Malaysia, 2019, pp. 1-5, doi: 10.1109/ICSCC.2019.8843683.

Jha, A. K., Singh, P. P. and Dwivedi, P. (2019). Maithili Text-to-Speech System, 2019 IEEE International Conference on Electronics, Computing and Communication Technologies (CONECCT), Bangalore, India, 2019, pp. 1-6, doi: 10.1109/CONECCT47791.2019.9012903.

Kardava, I., Antidze, J., & Gulua, N. (2016). Solving the problem of the accents for speech recognition systems. *International Journal of Signal Processing Systems*, *4*(3), 235-238.

Korde, V., & Mahender, C. N. (2012). Text Classification and Classifiers: A Survey. *International Journal of Artificial Intelligence & Applications*, *3*(2), 85–99.

Lodhi, H., Saunders, C., Shawe-Taylor, J., Cristianini, N., & Watkins, C. (2002). Text Classification using String Kernels. *Journal of Machine Learning Research*, *2*, 419–444. https://doi.org/10.1162/153244302760200687

Morwal, S., Jahan, N., & Chopra, D. (2012). Named entity recognition using hidden Markov model (HMM). *International Journal on Natural Language Computing (IJNLC)*, *1*(4), 15-23.

Mustafa MB., Ainon R.N. Emotional speech acoustic model for Malay: iterative versus isolated unit training, Journal of the Acoustical Society of America, 134(4), pp. 3057-3066, October, 2013.

Mustafa, M. B., Don, Z. M., Ainon, R. N., Zainuddin, R., & Knowles, G. (2014). Developing an HMM-based speech synthesis system for Malay: a comparison of iterative and isolated unit training. *IEICE TRANSACTIONS on Information and Systems*, *97*(5), 1273-1282.

Nthite, T. and Tsoeu, M. (2020). End-to-End Text-To-Speech synthesis for under resourced South African languages, 2020 International SAUPEC/RobMech/PRASA Conference, Cape Town, South Africa, 2020, pp. 1-6, doi: 10.1109/SAUPEC/RobMech/PRASA48453.2020.9041030.

Ranaivo-Malançon, B. (2006). Automatic identification of close languages-case study: Malay and Indonesian. *ECTI Transactions on Computer and Information Technology (ECTI-CIT)*, *2*(2), 126-134.

Rekha, M., Srinivas, S. Y., & Reddy, P. P. (2014). A Review on Pattern Discovery Techniques of Web Usage Mining. *Journal of Engineering Research and Applications Www.Ijera.Com ISSN*, *4*(4), 131–136.

Rohrer, B., Franks, L., & Ericson, G. (2016). How to choose machine learning algorithms | Microsoft Docs. Retrieved March 19, 2017, from https://docs.microsoft.com/en-us/azure/machine-learning/machine-learning-algorithm-choice

San-Segundo, R., Montero, J. M., Giurgiu, M., Muresan, I., & King, S. (2013). Multilingual Number Transcription for Text-to-Speech Conversion. *8th ISCA Workshop on Speech Synthesis (SSW)*, 65–69.

Shetake, P., Patil, A., & Jadhav, P. (2014). Review of text to speech conversion methods. *International Journal of Industrial Electronics and Electrical Engineering*, *2*(8), 29-35.

Sproat, R., Black, A. W., Chen, S., Kumar, S., Ostendorf, M., & Richards, C. (2001). Normalization of non-standard words. *Computer Speech and Language*, *15*, 287–333.

Swetha, N., & Anuradha, K. (2013). TEXT-TO-SPEECH CONVERSION. *International Journal of Advanced Trends in Computer Science and Engineering*, *2*(6), 269–278.

Tahir, M. A., Kittler, J., & Bouridane, A. (2016). Multi-label classification using stacked spectral kernel discriminant analysis. *Neurocomputing*, *171*, 127–137.

Tran, D. C., Ahamed Khan, M. K. A. and Sridevi, S. (2020). On the Training and Testing Data Preparation for End-to-End Text-to-Speech Application, 2020 11th IEEE Control and System Graduate Research Colloquium (ICSGRC), pp. 73-75, doi: 10.1109/ICSGRC49013.2020.9232605.

Trstenjak, B., Mikac, S., & Donko, D. (2014). KNN with TF-IDF based framework for text categorization. In *Procedia Engineering* (Vol. 69, pp. 1356–1364).